%% file: II2202-report.tex
\documentclass[12pt,twoside,english]{article}
\usepackage[utf8]{inputenc}
\usepackage{multicol} 

\input{lib/includes}
\input{lib/acronyms}
\input{lib/values}

\title{Efficient Implementation of CRYSTALS-KYBER Key Encapsulation Mechanism on ESP32}
\author{
    Fabian Segatz \\ \texttt{segatz@kth.se}
    \and 
    Muhammad Ihsan Al Hafiz \\ \texttt{miahafiz@kth.se} 
}
\date{December 2022}

\makeatletter
\let\ps@plain\ps@fancy 
\makeatother

\begin{document}

\maketitle

\begin{abstract}
\label{sec:abstract}

Kyber, an IND-CCA2-secure lattice-based post-quantum key-encapsulation mechanism, is the winner of the first post-quantum cryptography standardization  process of the US National Institute of Standards and Technology. In this work, we provide an efficient implementation of Kyber on ESP32, a very popular microcontroller for Internet of Things applications. We hand-partition the Kyber algorithm to enable utilization of the ESP32 dual-core architecture, which allows us to speed up its execution by \scenTwoKeygenSpeedUp (keygen), \scenTwoEncSpeedUp (encaps) and \scenTwoDecSpeedUp (decaps). We also explore the possibility of gaining further improvement by utilizing the ESP32's SHA and AES coprocessor and achieve a culminated speed-up of \scenThreeKeygenSpeedUp (keygen), \scenThreeEncSpeedUp (encaps) and \scenThreeDecSpeedUp (decaps)


\end{abstract}


\selectlanguage{english}
\tableofcontents

\clearpage

\section*{List of Acronyms and Abbreviations}
\label{list-of-acronyms-and-abbreviations}
\begin{multicols}{2}
\renewcommand{\glossarysection}[2][]{} 
\printglossary[type=\acronymtype,nonumberlist]
\end{multicols}




\section{Introduction}
\label{sec:introduction}

Due to the rapidly growing market for \gls{IOT} applications, the ESP platform series from Espressif System increased its market share in the past years. In December 2017, Espressif System shipped 100-Million devices of the ESP series across the world \cite{espressif_systems_shanghai_co_ltd_milestones_nodate}. To secure the communication channel in \gls{IOT} applications, Anand et al. \cite{anand_2019} implemented \gls{RSA} encryption with a public key concept in ESP32, one of the microcontrollers from the ESP series. \gls{RSA} encryption was widely used in previous \gls{IOT} applications \cite{nandanavanam_2022}. Unfortunately, several studies \cite{buchmann_2017,ziying_2022,schoffel_secure_2022} stated that Shor's polynomial-time algorithm \cite{shor_1999} will be able to break established cryptography security, including \gls{RSA} encryption, once quantum computers are available. In 2016, the \gls{NIST} released a formal global call for standardizing new \gls{PKE} methods and \glspl{KEM} suitable for \gls{PQC} \cite{ziying_2022,takagi2016post}. In July 2022, Kyber was announced as the first post-quantum \gls{KEM} algorithm standard \cite{alagic_2022} after a six-year selection process. There are open-source Kyber implementations existing for the ARM Cortex M4 \cite{Kannwischer_2022, botros_2019} and the Haswell processor architecture \cite{bos_2022}. 

This work aims to find an efficient implementation of \gls{CRYSTALS}-Kyber on ESP32 that shows the advantage of utilizing the ESP32's dual-core architecture, as well as \gls{AES} and \gls{SHA}. We plan to give an answer on how much the execution time of Kyber’s keypair-generation, encapsulation and decapsulation algorithm can be reduced by utilizing the ESP32's hardware features. The purpose of an efficient implementation of Kyber on ESP32 is to enable the usage of \gls{PQC} in many \gls{IOT} applications and to encourage handling the future world's security requirements ahead of time. It is expected that utilizing both cores of the ESP32 by paralleling parts of the Kyber algorithm and utilizing other available hardware accelerators will result in a more time-efficient execution compared to using a simple single-core implementation.
To the best of our knowledge, this is the first published optimized implementation of Kyber \gls{KEM} on ESP32. 

\subsection*{Contribution}

\begin{enumerate}
    \item The ESP32 processor consists of two Harvard Architecture Xtensa LX6 CPUs. When utilizing the \gls{ESP-IDF} framework and its built-in version of FreeRTOS, it is possible to run threads on the respective cores. However, to speed up the execution time of a program, it must be partitioned into tasks and rearranged in a way that they can be executed in parallel in individual threads. This work presents an effective scheme of parallelizing the Kyber's \gls{IND-CPA} functionalities.
    \item This paper presents results on how much speed-up is achievable by utilizing the ESP32's hardware accelerators for \gls{SHA} and \gls{AES} when executing the 90s variant of Kyber.
\end{enumerate}

\subsection*{Organization}

In Section \ref{sec:background}, we aim to briefly describe Kyber \gls{KEM}, the target platform and the tools that we used. In Section \ref{sec:methods}, we describe the methods for measuring the performance of our implementations. Further, we elaborate on how to utilize the hardware accelerators and explain task parallelism as a basic principle for our dual-core implementation. In Section \ref{sec:implementation}, the implementations will be presented. This is followed by the results in Section \ref{sec:results} and the discussion in Section \ref{sec:discussion}. Finally, we want to conclude our work in Section \ref{sec:conclusion} and elaborate on future research.

\section{Background}
\label{sec:background}

This section gives a brief overview of the Kyber algorithm, the used hardware platform ESP32, and the tools we used in this research. 


\subsection{CRYSTALS-Kyber}
\label{sec:kyber}

Kyber is a lattice-based \gls{KEM} that was selected as the \gls{PQC} standard in July 2022 \cite{alagic_2022}. Its security is based on the \gls{MLWE} \cite{langlois_2012}. Kyber consists of an \gls{IND-CPA}-secure \gls{PKE} that is used for the \gls{IND-CCA2}-secure \gls{KEM}. The complete algorithm is specified by Avanzi et al. \cite{avanzi_2021}.

\subsubsection*{Kyber PKE}
\label{sec:kyber_pke}

\Gls{PKE} was first introduced by Diffie and Hellman in 1976 \cite{diffie_1976}. It allows asymmetric cryptography utilizing a pair of keys. When one of those keys gets used for encryption, the other one is needed for decryption. This allows one of the keys to be published (public key) so that anyone can use it, while the other remains private (secret key).

Every \gls{PKE} scheme relies on some computationally hard mathematical problem. Kyber is a lattice-based cryptography algorithm and relies on the \gls{MLWE}. In this work, we want to give a high-level understanding of the data flow because this is what is relevant for the later implementation. Avanzi et al. published a written specification of Kyber \cite{avanzi_2021} that defines all the primitives of the following visualization.

\begin{figure}[]
    \centering
    \includegraphics[width=0.6\textwidth]{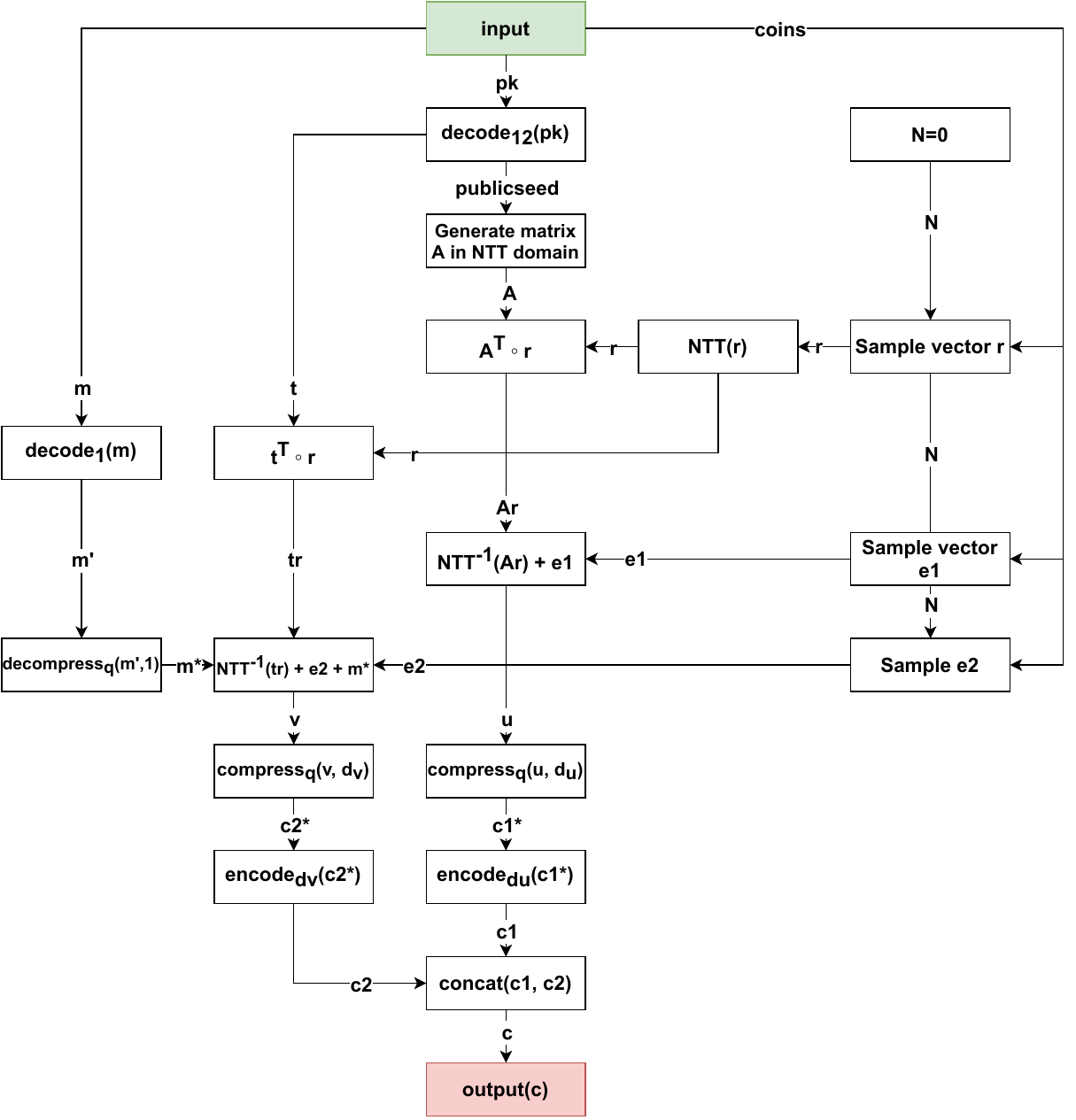}
    \caption{Visualization of the \gls{IND-CPA}-secure encryption algorithm indcpa\_enc()}
    \label{fig:kyber_pke_enc}
\end{figure}

\begin{figure}[]
    \centering
    \begin{minipage}[t]{0.45\textwidth}
        \centering
        \includegraphics[width=\textwidth]{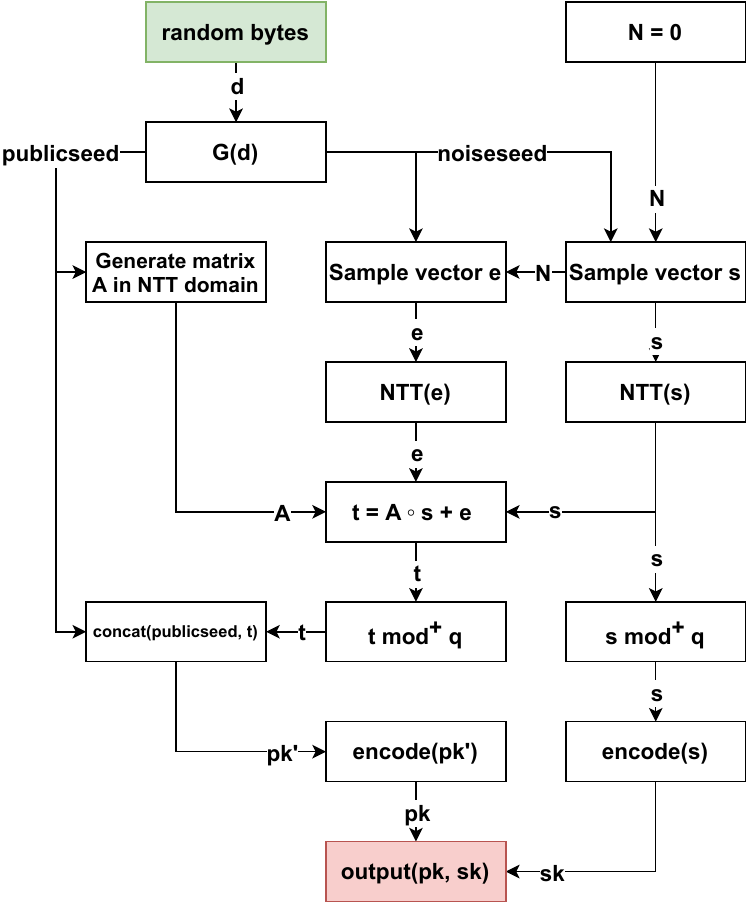}
        \caption{Visualization of the \gls{IND-CPA}-secure keypair generation algorithm indcpa\_keygen()}
        \label{fig:kyber_pke_keygen}
    \end{minipage}\hfill
    \begin{minipage}[t]{0.45\textwidth}
        \centering
        \includegraphics[width=\textwidth]{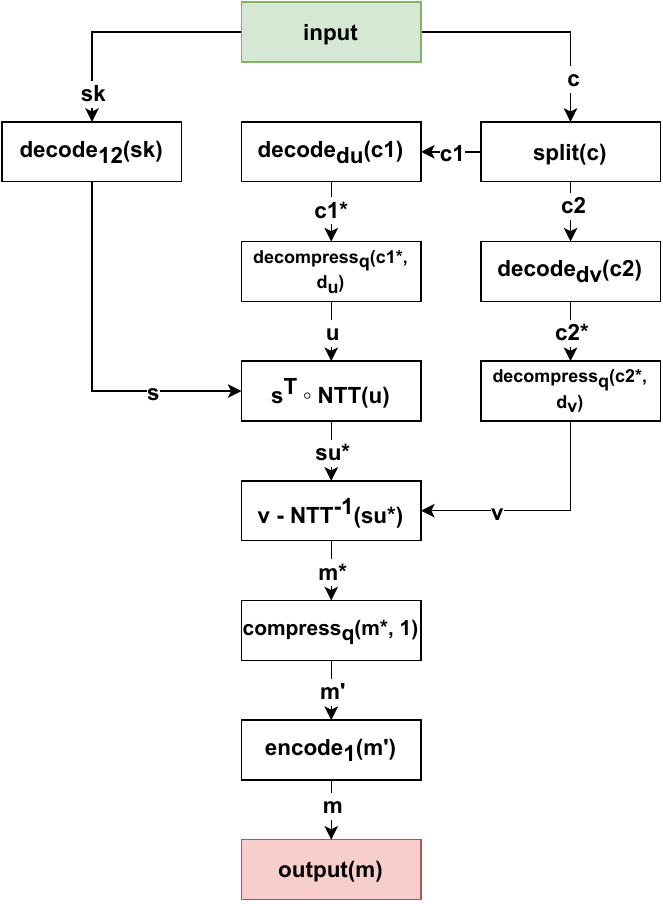}
        \caption{Visualization of the \gls{IND-CPA}-secure decryption algorithm indcpa\_dec()}
        \label{fig:kyber_pke_dec}
    \end{minipage}\hfill
\end{figure}

Figure \ref{fig:kyber_pke_keygen} shows the data flow of the keypair generation. The entry point, marked in green, shows that some random bytes must first be created. From those, we obtain a public- and a noise seed. The \texttt{publicseed} is used to create a matrix \texttt{A}, while the \texttt{noiseseed} is used to sample the vectors \texttt{e} and \texttt{s}. All of those variables need to be in the \gls{NTT}-space, a special version of a discrete Fourier transform. Because of the way that Kyber is designed, the required matrix multiplication ($A \cdot s$) consists of additions, if it is executed in the \gls{NTT}-space \cite{avanzi_2021}. Ultimately, the value \texttt{t} and the \texttt{publicseed} get concatenated and encoded into the public key \texttt{pk}, while the vector \texttt{s} gets concatenated into the secret key \texttt{sk}.

Figure \ref{fig:kyber_pke_enc} presents the encryption data flow. There must be three inputs: the public key \texttt{pk}, the message text \texttt{m} and some random bytes \texttt{coins}. From the coins, the error-vectors \texttt{r} and \texttt{e1} as well as the error-value \texttt{e2}. The errors are used together with \texttt{pk} to generate some vector \texttt{u}. The message text \texttt{m} gets encrypted with some parts of the process that leads to \texttt{u} and ultimately results in vector \texttt{v}. Those two values get compressed and encoded before they get concatenated to the final ciphertext \texttt{c}.

Figure \ref{fig:kyber_pke_dec} visualizes the decryption method. The input parameters are the secret key \texttt{sk} and the ciphertext \texttt{c}. The ciphertext gets parted, decoded and decompressed into the vectors \texttt{u} and \texttt{v}. 
Those get then used together with the secret key \texttt{sk} to recreate the message text \texttt{m}.

\subsubsection*{Kyber KEM}
\label{sec:kyber_kem}

Shoop proposed the first \gls{KEM} in 2000 \cite{shoup_2000}. He defines it as a special use-case of \gls{PKE} where the encryption algorithm takes no other input than the recipient's public key. Instead of a message text, it automatically generates a random bit string of a specified length. The encrypted bit string gets transferred to the recipient that can now decrypt it using their secret key.

Only the recipient and sender now know this random bit string, which is also called a shared secret. Instead of using a computationally extensive asynchronous cryptography scheme, the party can now use the shared secret for a symmetric cryptography scheme such as \gls{AES}.

\begin{figure}[h]
    \centering
    \begin{minipage}[t]{0.3\textwidth}
        \centering
        \includegraphics[width=\textwidth]{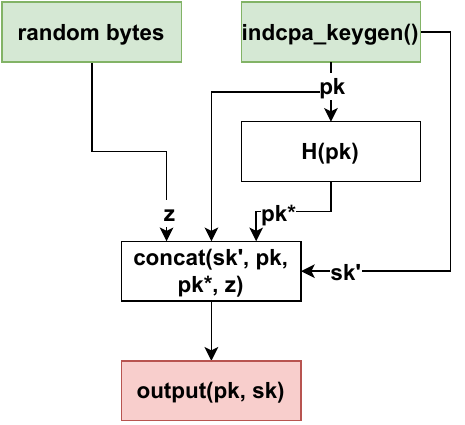}
        \caption{Visualization of the \gls{IND-CCA2}-secure keypair generation algorithm kem\_keygen()}
        \label{fig:kyber_kem_keygen}
    \end{minipage}\hfill
    \begin{minipage}[t]{0.3\textwidth}
        \centering
        \includegraphics[width=\textwidth]{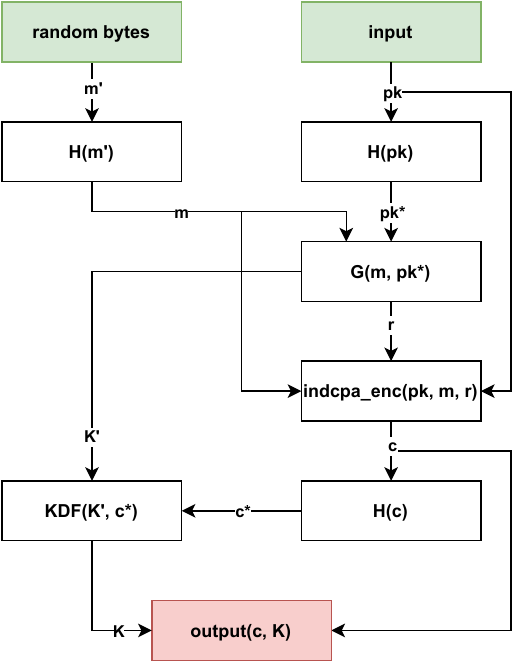}
        \caption{Visualization of the \gls{IND-CCA2}-secure encapsulation algorithm kem\_encaps()}
        \label{fig:kyber_kem_encaps}
    \end{minipage}\hfill
    \begin{minipage}[t]{0.3\textwidth}
        \centering
        \includegraphics[width=\textwidth]{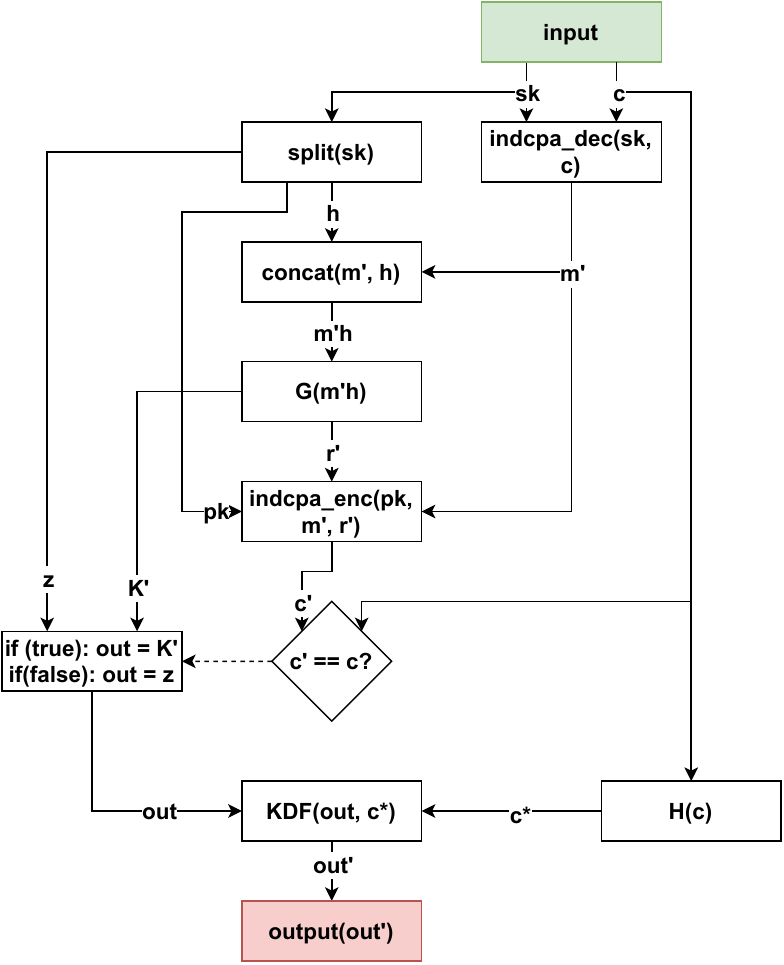}
        \caption{Visualization of the \gls{IND-CCA2}-secure decapsulation algorithm kem\_decaps()}
        \label{fig:kyber_kem_decaps}
    \end{minipage}\hfill
\end{figure}

The authors of Kyber decided to use a slightly tweaked Fujisaki–Okamoto transform \cite{fujisaki_1999} to construct a \gls{IND-CCA2}-secure \gls{KEM}. The Figures \ref{fig:kyber_kem_keygen}, \ref{fig:kyber_kem_encaps}, \ref{fig:kyber_kem_decaps} visualize the data flow.

\subsubsection*{Kyber Parameters}

The written specification defines the three parameter sets KYBER512, KYBER768 and KYBER1024, where KYBER512 is the most efficient and Kyber1024 is the most secure. Further, two different implementation variants exist, the standard and the 90s variant. Kyber needs an \gls{XOF}, two hash functions, a \gls{PRF}, and a \gls{KDF}. In the standard variant, the authors of Kyber decided to rely on only one underlying primitive for all those functions, namely the SHAKE functions based on Keccak \cite{bertoni_2011}. This was done to reduce code size and exploitable weaknesses. In the 90s variant, however, the approach relies entirely on primitives,  namely AES, SHA-256 and SHA-512, that are already standardized for a long time and accelerated in hardware on a large variety of platforms. \cite{avanzi_2021}.

\subsection{Platform}
\label{sec:platform}

ESP32 is a low-cost and high-performance microcontroller that is manufactured by Espressif System. Its built-in WiFi and Bluetooth are why it has become so popular for \gls{IOT} applications. The ESP32 is a model belonging to the ESP series of microcontrollers, which was sold over 100 million times until the end of 2017\cite{espressif_systems_shanghai_co_ltd_milestones_nodate}. The ESP32 is a dual-core system with two Xtensa 32-bit \glspl{CPU} and a five-stage pipeline that can be operated at up to 240 MHz. The hardware architecture features several co-processors to accelerate popular algorithms such as \gls{RNG}, \gls{AES}-128/256, \gls{SHA} and \gls{RSA} \cite{espressif_systems_esp32_2022}. The dual-core system on ESP32 can become an advantage for processing a high amount of data. Wang et al. \cite{wang_saber_2020} showed that exploiting the dual-core on ESP32 could speed up the implementation of Saber, which is a \gls{CCA}-secure lattice-based post-quantum \gls{KEM} that is similar to Kyber.

\subsection{Tools}
\label{sec:tools}

The \gls{ESP-IDF} is a common framework for developing applications for ESP32 \cite{Maier_2017}. It includes a tool-chain and an extensive \gls{API}. The \gls{ESP-IDF} facilitates  a modular programming approach with components. One of them is FreeRTOS, a popular \gls{RTOS} for embedded devices, which allows to create threads, that can be pinned to either of the two ESP32 cores \cite{espressif_freertos_2022}.
Another component is the open-source C library MbedTLS, which implements cryptographic primitives \cite{arm_limited_tls_2022}. It is included into the \gls{ESP-IDF}, as a wrapper for accessing the low-level register to utilize the accelerator functionalities of the ESP32.


\section{Methods}
\label{sec:methods}

This research applies an empirical research method \cite{peter_bock_getting_2001}, as our goal is to present a possible efficient performance improvement for Kyber on ESP32 and not to find the best-performing implementation. The empirical research will be conducted and analyzed using a quantitative method. We will gather numerical data from several implementation scenarios of Kyber on ESP32. Those scenarios are explained in Section \ref{sec:implementation}. The empirical evidence from the experiment results will be used to generate a concrete conclusion from this research. 


\subsection{Clock tick cycle measurement}
\label{sec:clocktick}

The goal of this work is to find an efficient implementation of Kyber on ESP32. To our knowledge, there are three metrics for efficiency that are relevant for embedded systems: Power consumption, memory usage and execution speed. We will measure the clock tick cycle count of the processor as a metric of the execution speed. 

A \gls{CPU} operates by executing the instruction in assembly language that is coordinated by clock. The sequential high and low signal in a clock is called a cycle or clock tick cycle \cite{howard_gilbert_clocks_2010}. Every single line of operation like addition or moving data requires a certain amount of clock tick cycles to complete the process. The amount of tick cycles for the operation depends on the assembly code that is generated by the compiler and the architecture of a digital system. In other words, the clock tick cycle count is related to the instruction set of a processor, the way that the compiler translates a program into those instructions and the computational effort of the program itself.

The clock tick cycle count measurement was used to evaluate the performance of each implementation step. Ultimately, we used it to measure the three functions of the Kyber \gls{KEM} process to evaluate the total speedup. The clock tick count was used to evaluate the speed of the process because it can be analyzed for several clock configurations. Wang et al. \cite{wang_saber_2020}, and Heinz et al. \cite{heinz2022} measured the clock tick count to benchmark their cryptography algorithm implementations. Therefore, we used the clock tick count from ESP32 to benchmark the Kyber implementation in several scenarios. The clock tick counting was acquired by utilizing the official function in the ESP32 library \cite{espressif_web_idf_2022} which is \verb|esp_cpu_get_cycle_count(void)|. The function returned a 32-bit unsigned integer of the current cycle count of the \gls{CPU}. We called this function once, just before the assessed code segment and once after. The difference of those two measurements reflects then on the total clock tick cycle count. From the clock tick cycle count, we can then calculate the execution time (runtime) of the code segment under test, by dividing it through the clock frequency of the \gls{CPU} (see Equation \ref{eq:runtime}). The same calculation approach was used also by Wang et al. \cite{wang_saber_2020} to get their run time.

\begin{equation}
\label{eq:runtime}
Run time (ms) = \frac{Cycle Count}{frequency}*1000
\end{equation}

\subsection{Task parallelism}
\label{sec:parallel}

In the scope of this work, a task $T$ is understood as a junk of computational work. A program $P$, which is a task $T$ itself, can be partitioned into many smaller tasks so that $P=\{T_0, T_1, ..., T_n\}$. Tasks are dependent on each other - most tasks have predecessors $P$ that must be executed before $T$.

In Section \ref{sec:kyber}, we presents the Figures \ref{fig:kyber_pke_keygen}, \ref{fig:kyber_pke_enc}, \ref{fig:kyber_pke_dec} to show the data flow within Kyber's \gls{PKE} algorithms. From the graphs, it can be seen, that many of the tasks could be executed in parallel as they are not directly dependent on each other. 

Whether this task parallelization results in an execution time speed up is mainly dependent on the computational cost, versus the communication cost, that is introduced through moving data between individual \glspl{CPU} \cite{barton_2011}.

When parallelizing, it is important to prevent race conditions, where one \gls{CPU} already tries to compute a task $T$, whose predecessor task $P$ is assigned to another \gls{CPU}, but not yet executed. Semaphores can be used to introduce synchronisation points, which help to resolve this issue in a simple but effective way. FreeRTOS offers an off-the-shelf semaphore implementation \cite{freertos_semaphore_2022}.



\section{Implementation}
\label{sec:implementation}

In this study, we evaluated Kyber512 in the 90s variant, since it presents the best capability for efficiency improvement on the ESP32. Kyber512 is the smallest parameter set and thus needs the least computational effort. The 90s variant relies only on AES and SHA-256 primitives of which both can be computed with respective acceleration hardware that is available on the ESP32.



We split up our implementation efforts into 3 scenarios. In this section, we provide a brief overview of those. The source code is available on GitHub \cite{segatz_2022}. \\


\noindent \textbf{Scenario 1: Kyber512-90s using a single-core.} As the baseline for this study, the C-implementation by Bos et al. \cite{bos_2022} was ported to the ESP32 architecture. The source code for the different primitives was transformed into a component structure as suggested by the \gls{ESP-IDF} build system \cite{espressif_web_buildsystem_2022}. The function \texttt{void app\_main(void)} has to be created as a program entry point \cite{espressif_web_startup_2022}. Its only task is to initializes one FreeRTOS thread that is pinned to one of the cores via the function \texttt{BaseType\_t xTaskCreatePinnedToCore(TaskFunction\_t pvTaskCode, ...)}, where \texttt{pvTaskCode} is the handle for the test thread. The test thread encapsulates the three \gls{KEM} functions: Keypair generation, encryption and decryption. Only one function of the implementation by Bos et al. \cite{bos_2022} has to be replaced to make the code compile and run on the ESP32. Namely, the \gls{RNG} functionality, which previously used a Linux system call and has to be replaced. Instead the \gls{ESP-IDF}'s function \texttt{void esp\_fill\_random(void *buf, size\_t len)} was used, which fills the memory pointer \texttt{buf} with \texttt{len} random bytes. \\


\noindent \textbf{Scenario 2: Kyber512-90s using the  dual-core.} The reason for parallelizing programs is to reduce their execution time. However, the way how to achieve this goal is highly dependent on the program and the execution platform themselves.  \cite{katagiri_2019}. Specifically, sequential data flow in the program and the communication time between the \gls{CPU} cores and the memory must be regarded.

In Section \ref{sec:parallel}, we stated that Kyber's \gls{PKE} algorithms have good parallelization capabilities. We applied an empirical approach, which focused on keeping the data transfer between the two cores of the ESP32 as low as possible. The main focus was on organising the schedule such as, that sequential data processing would still be done on the same core. Ultimately, we came up with the following resource allocation plan, and then we expect to perform the best.

\begin{figure}[H]
    \centering
    \includegraphics[width=\textwidth]{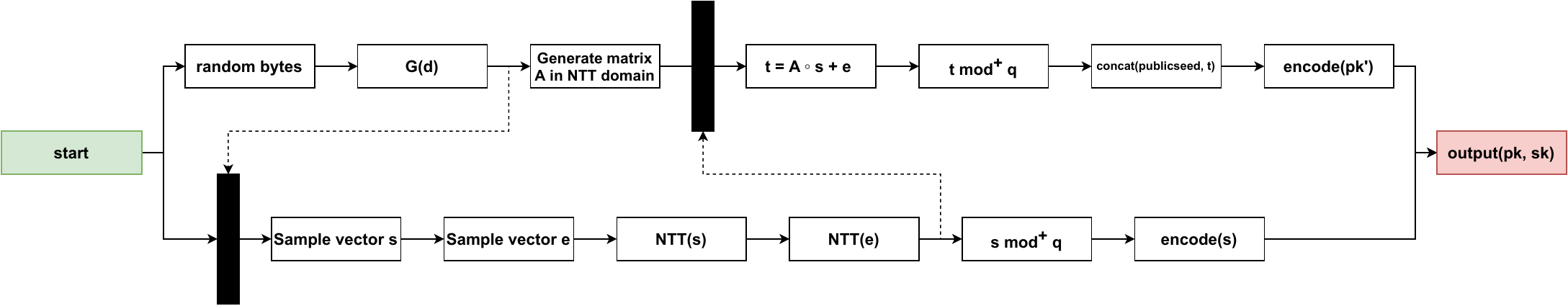}
    \caption{Execution scheme for dual-core keypair generation algorithm}
    \label{fig:indcpa_keygen_dual}
\end{figure}

\begin{figure}[H]
    \centering
    \includegraphics[width=\textwidth]{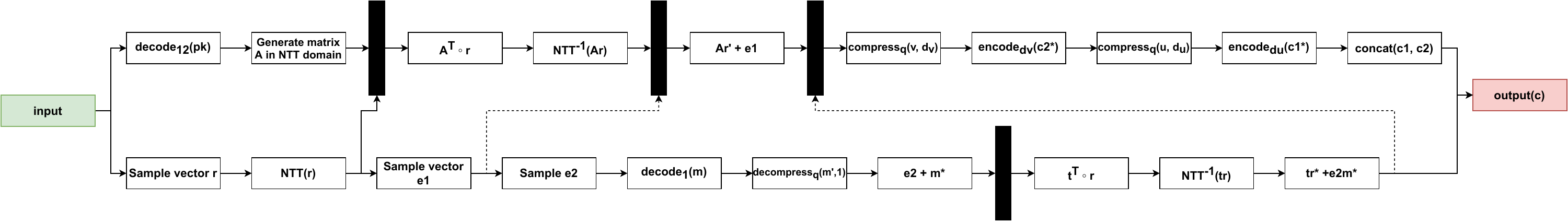}
    \caption{Execution scheme for dual-core encryption algorithm}
    \label{fig:indcpa_enc_dual}
\end{figure}

\begin{figure}[H]
    \centering
    \includegraphics[width=\textwidth]{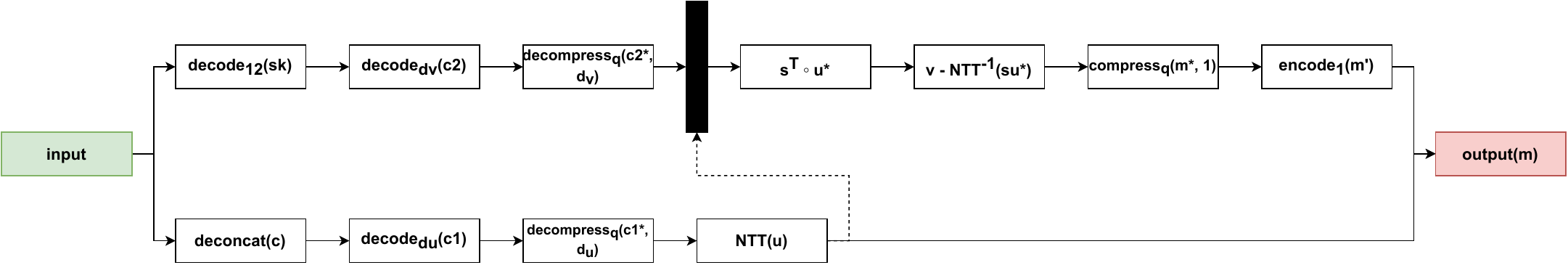}
    \caption{Execution scheme for dual-core decryption algorithm}
    \label{fig:indcpa_dec_dual}
\end{figure}

The Figures \ref{fig:indcpa_keygen_dual}, \ref{fig:indcpa_enc_dual}, and \ref{fig:indcpa_dec_dual} visualize the implemented resource allocation. Tasks are presented as white rectangular boxes. Each row of tasks is executed by one \gls{CPU}-core. The black boxes represent the synchronization points that are introduced by semaphores. When a core reaches the synchronization point, it has to wait until the semaphore becomes free. This is triggered when the other core reaches the execution state where the dashed arrow originates from. \\


\noindent \textbf{Scenario 3: Kyber512-90s using the dual-core and hardware accelerators.}  \gls{SHA} and  \gls{AES} accelerators with MbedTLS library support were used in scenario 3. The \gls{SHA}-256 and the \gls{SHA}-512 were used for hashing process because the Kyber 90s variant uses those algorithms as standard algorithms \cite{avanzi_2021}. The MbedTLS provides support for accessing the \gls{SHA} and \gls{AES} accelerator for ESP32 and is commonly used as hashing process for cryptography algorithms. Bürstinghaus-Steinbach et al. \cite{burstinghaus_2020} utilized MbedTLS in their implementation of Kyber in ESP32. For the implementation in this paper, the hashing function in C code that has been developed by Bos et al. \cite{bos_2022} was substituted with the equivalent function from MbedTLS. The \gls{SHA}-256 uses the function \texttt{mbedtls\_sha256(IN, INBYTES, OUT, 0)}, and the \gls{SHA}-512 uses the function \texttt{mbedtls\_sha512(IN, INBYTES, OUT, 0)}. The function input for those \gls{SHA} processes are similar, where \texttt{IN} is the input data, \texttt{INBYTES} is the number of input bytes data, and \texttt{OUT} is the output hash. The zero number as the last input function for  \gls{SHA}-256 indicates the function is not  \gls{SHA}-224, and the zero number as the last input function for  \gls{SHA}-512 indicates the function is not  \gls{SHA}-384. For the \gls{AES} accelerator, we used it at the pseudo-random function in the \gls{AES} hashing process in the function of \texttt{kyber\_aes256ctr\_prf}. For activating the accelerator, we substituted the original \gls{AES} function from Bos et al. \cite{bos_2022} with the equivalent function from the MbedTLS library. In addition, the verification was conducted by comparing the key result from the encryption process with the key result from the decryption process.

\section{Results}
\label{sec:results}

The implementation is written in C language and is based on the \gls{ESP-IDF} development framework in version 5.0. The compilation is done by the framework's default compiler, which is GCC in version 8.4.0. We did not activate any compiler optimization. The execution was performed on an ESP32-S3-DevKitC-1 development board \cite{esp32_devboards} with a clock frequency of 160 MHz. The clock frequency is only relevant for the run time calculation. The compiler settings were kept default as they were specified when setting up the project with the \gls{ESP-IDF}.

Firstly, we evaluated the performance of our dual-core implementation. Table \ref{tab:results_dualcore} shows the results of our measurements for the clock-tick cycle of the Kyber \gls{PKE} functions. We compared and calculated a speedup ratio for the dual-core implementation against the single-core implementation. The clock tick cycle measurement was conducted in an isolated way, so there was no interference with other functions to guarantee the credibility of measurement from each function. Based on Table \ref{tab:results_dualcore}, the function of \gls{IND-CPA} for decryption shows more execution cycle count in dual-core compared to the single-core implementation, which is inversely proportional to the other functions.

\begin{table}[h]
    \centering
    \caption[Performance of IND-CPA]{Performance of \gls{IND-CPA} function in single and dual core implementation}
    \begin{tabular}{c|c|ccc}

     Algorithm & Implementation & Cycle count & Run time (ms) & Speedup ratio \\ \hline\hline
    \multirow{2}{*}{\gls{IND-CPA} key-pair generation} & Single-core & \indcpaKeygenSingleCycle & \indcpaKeygenSingleRunTime & \indcpaKeygenSingleSpeedUp \\
    & Dual-core & \indcpaKeygenDualCycle & \indcpaKeygenDualRunTime & \indcpaKeygenDualSpeedUp \\
    \hline
    \multirow{2}{*}{\gls{IND-CPA} encryption } & 
    Single-core & \indcpaEncSingleCycle & \indcpaEncSingleRunTime & \indcpaEncSingleSpeedUp \\
    & Dual-core & \indcpaEncDualCycle & \indcpaEncDualRunTime & \indcpaEncDualSpeedUp \\
    \hline
    \multirow{2}{*}{\gls{IND-CPA} decryption} & 
    Single-core & \indcpaDecSingleCycle & \indcpaDecSingleRunTime & \indcpaDecSingleSpeedUp \\
    & Dual-core & \indcpaDecDualCycle & \indcpaDecDualRunTime & \indcpaDecDualSpeedUp \\
    \hline\hline
\end{tabular}
    \label{tab:results_dualcore}
\end{table}



Secondly, we evaluated the performance speedup using the \gls{SHA} and \gls{AES} accelerator. We measured the \gls{SHA}-256, \gls{SHA}-512, and \gls{AES} primitives, with and without utilizing the accelerator. Table \ref{tab:results_sha} presents the measurements. It is observable that using the accelerator reduces the clock tick cycle count in both cases.

\begin{table}[h]
    \centering
    \caption[Performance of \gls{SHA}]{Performance of \gls{SHA} and \gls{AES} function with and without accelerator}
    \begin{tabular}{c|c|ccc}
    Algorithm & Implementation & Cycle count & Run time (ms) & Speedup ratio \\ \hline\hline
    \multirow{2}{*}{\gls{SHA}-256} & 
    Without accelerator & \shaTfsNoAccCycle & \shaTfsNoAccRunTime & \shaTfsNoAccSpeedUp \\
    & With accelerator & \shaTfsAccCycle & \shaTfsAccRunTime & \shaTfsAccSpeedUp \\
    \hline
    \multirow{2}{*}{\gls{SHA}-512} & 
    Without accelerator & \shaFotNoAccCycle & \shaFotNoAccRunTime & \shaFotNoAccSpeedUp \\
    & With accelerator & \shaFotAccCycle & \shaFotAccRunTime & \shaFotAccSpeedUp \\
    \hline
    \multirow{2}{*}{\gls{AES}} & 
    Without accelerator & \aesNoAccCycle & \aesNoAccRunTime & \aesNoAccSpeedUp \\
    & With accelerator & \aesAccCycle & \aesAccRunTime & \aesAccSpeedUp \\
    \hline\hline
\end{tabular}
    \label{tab:results_sha}
\end{table}

Finally, we measured the effective speed-up our implementation efforts can achieve on the Kyber \gls{KEM} functionality. Scenario 1 is the original implementation by Bos et al. \cite{bos_2022} ported to the ESP32 platform and is used as a reference for the comparison against our optimizations. Scenario 2 uses dual-core implementations for \gls{PKE} keypair generation and encryption but not for decryption, as this has been shown to be slowing down the execution, as presented in Table \ref{tab:results_dualcore}. Scenario 3 uses the \gls{SHA} accelerator on top of Scenario 2 and reflects there for our final proposition for an efficient implementation of \gls{CRYSTALS}-Kyber \gls{KEM} on ESP32. 

\begin{table}[h]
    \centering
    \caption[Performance of Kyber512-90s on ESP32]{Performance of Kyber512-90s on ESP32 
    }
    \begin{tabular}{c|cccc}
    Implementation & Algorithm & Cycle count & Run time (ms) & Speedup ratio \\ \hline\hline
    \multirow{3}{*}{Scenario 1 \cite{bos_2022}} & Key Generation & \scenOneKeygenCycle & \scenOneKeygenRunTime & \scenOneKeygenSpeedUp \\
    & Encapsulation & \scenOneEncCycle & \scenOneEncRunTime & \scenOneEncSpeedUp \\
    & Decapsulation & \scenOneDecCycle & \scenOneDecRunTime & \scenOneDecSpeedUp \\
    \hline
    \multirow{3}{*}{Scenario 2} & Key Generation & \scenTwoKeygenCycle & \scenTwoKeygenRunTime & \scenTwoKeygenSpeedUp \\
    & Encapsulation & \scenTwoEncCycle & \scenTwoEncRunTime & \scenTwoEncSpeedUp \\
    & Decapsulation & \scenTwoDecCycle & \scenTwoDecRunTime & \scenTwoDecSpeedUp \\
    \hline
    \multirow{3}{*}{Scenario 3} & Key Generation & \scenThreeKeygenCycle & \scenThreeKeygenRunTime & \scenThreeKeygenSpeedUp \\
    & Encapsulation & \scenThreeEncCycle & \scenThreeEncRunTime & \scenThreeEncSpeedUp \\
    & Decapsulation & \scenThreeDecCycle & \scenThreeDecRunTime & \scenThreeDecSpeedUp \\
    \hline
\end{tabular}
    \label{tab:results_final}
\end{table}

%

\section{Discussion}
\label{sec:discussion}

The dual-core parallelization does not always guarantee the process will have more speed up execution. 
The reason for speed down in decryption process in Table \ref{tab:results_dualcore} is that the parallelization effort shown in Figure \ref{fig:indcpa_dec_dual} does not significantly contribute to the total execution cycle count compared to the process after the parallelization. The parallelization effort requires additional processes such as defining the task in dual-core, defining the semaphores, semaphores counting, and other task management process. Those additional processes in the \gls{IND-CPA} for decryption take more cycle count than the reduction from parallelization that has been made. 
In addition, the \gls{SHA} accelerator with the MbedTLS library for ESP32 provides a significant speed-up for hashing process compared to the original \gls{SHA} function implementation in the Kyber from Bos et al. \cite{bos_2022}. 


Overall, the utilization of dual-core capability and hardware accelerator in ESP32 can accelerate the implementation of the Kyber algorithm in the process of key generation, encryption, and decryption, compared to single-core implementation without a hardware accelerator. The utilization of dual-core capability does not contribute significantly compared to the utilization of hardware accelerators. It is because the functions in the Kyber algorithm are mainly executed in sequential patterns with dependent variables from the previous execution. The cycle count reduction in Scenario 2 on Table \ref{tab:results_final} is because of the process parallelization \gls{IND-CPA} function in key generation and encryption. The decryption process in Scenario 2 has a reduction cycle count because it uses the \gls{IND-CPA} encryption process in the decryption process. Moreover, the hardware accelerator shows a significant reduction in cycle count in all processes because every process contains the hashing operation with SHA-256 and SHA-512.

To the best of our knowledge, there are no publications about the implementation of Kyber-512-90s on ESP32 with the benchmark in a clock cycle and run time. Kyber implementation at the microcontroller level was accomplished with ARM Cortex M4 microcontroller that has been reported in submission round 3 for the third round of the \gls{NIST} \gls{PQC} Standardization Process \cite{avanzi_2021}. 
In addition, Bürstinghaus-Steinbach et al. \cite{burstinghaus_2020} presented Kyber-512 implementation in ESP32 and got the run time of 12ms, 16ms, and 18ms for key-generation, encryption, and decryption, respectively. Therefore, since there is no exact implementation of Kyber-512-90s available in a similar platform, there is no chance of performing a fair comparison. The report on this paper can become a benchmark for the future implementation of the Kyber-512-90s variant in ESP32.

\section{Conclusion}
\label{sec:conclusion}


In 2016, the \gls{NIST} released a global call for standardizing \gls{PQC}. In July 2022, Kyber was chosen as the first standard for post-quantum \gls{KEM}. In this paper, we presented an efficient implementation of the Kyber-512 in the 90s variant on ESP32, which is one of the most common \gls{IOT} microcontrollers. The ESP32 features a dual-core \gls{CPU} architecture as well as a \gls{SHA} and \gls{AES} coprocessor, that we made use of to accelerate the Kyber \gls{KEM} functionalities: key-pair generation, encapsulation and decapsulation.
We presented an effective way of partitioning and parallelizing the Kyber \gls{PKE} key-pair generation function (\indcpaKeygenDualSpeedUp speed-up) and encryption function (\indcpaEncDualSpeedUp speed-up). This resulted in a speed-up of \scenTwoKeygenSpeedUp for the Kyber \gls{KEM} key-pair generation, \scenTwoEncSpeedUp for the encapsulation and \scenTwoDecSpeedUp for the decapsulation.
We replaced the original software implementation of \gls{AES} and \gls{SHA}, as presented in the official C implementation of Kyber by Bos et al. \cite{bos_2022}, with respective functions from the MbedTLS library to make use of the relevant hardware accelerators. When using those, we achieved an accumulated speed-up of \scenThreeKeygenSpeedUp for the key-pair generation, \scenThreeEncSpeedUp for the encapsulation and \scenThreeDecSpeedUp for the decapsulation.
The speed-up metrics were obtained by comparing the clock tick cycle count.




Future work can be conducted on exploring further optimizations. The generation of matrix A, which is necessary for both the \gls{KEM} key-pair generation and encapsulation, is by far the heaviest task with ca. 900.000 clock tick cycles, and it would make sense to compute half of the matrix on each core. Another option is to improve the parallelization schemes by calculating an optimized schedule after calculating the worst-case execution time of each task.

\bibliography{refs}
\bibliographystyle{myIEEEtran}

\end{document}

%% file: lib/includes.tex
\usepackage[paper=a4paper,dvips,top=1.5cm,left=1.5cm,right=1.5cm,
    foot=1cm,bottom=1.5cm]{geometry}

\usepackage{todonotes}          

\usepackage{mathptmx}
\usepackage[scaled=.90]{helvet}
\usepackage{courier}
\usepackage{bookmark}

\usepackage{fancyhdr}
\usepackage[utf8]{inputenc} 
\usepackage[english]{babel}
\usepackage{array}			 
\usepackage{graphicx}	                 
\usepackage{float}			 
\usepackage{color}                       
\usepackage{mdwlist}
\usepackage{tabularx}		         
\usepackage{multirow}	                 
\usepackage{dcolumn}	                 
\usepackage{url}	                 
\usepackage[perpage,para,symbol]{footmisc} 
\usepackage{siunitx} 

\usepackage[all]{hypcap}	 

\definecolor{darkblue}{rgb}{0.0,0.0,0.3} 
\definecolor{darkred}{rgb}{0.4,0.0,0.0}
\definecolor{red}{rgb}{0.7,0.0,0.0}
\definecolor{lightgrey}{rgb}{0.8,0.8,0.8} 
\definecolor{grey}{rgb}{0.6,0.6,0.6}
\definecolor{darkgrey}{rgb}{0.4,0.4,0.4}
\makeatletter
\renewcommand\paragraph{\@startsection{paragraph}{4}{\z@}%
  {-3.25ex\@plus -1ex \@minus -.2ex}%
  {1.5ex \@plus .2ex}%
  {\normalfont\normalsize\bfseries}}
\makeatother

\makeatletter
\renewcommand\subparagraph{\@startsection{subparagraph}{5}{\z@}%
  {-3.25ex\@plus -1ex \@minus -.2ex}%
  {1.5ex \@plus .2ex}%
  {\normalfont\normalsize\bfseries}}
\makeatother

\usepackage[acronym, nopostdot]{glossaries}
\glsdisablehyper
\makeglossaries


%% file: lib/acronyms.tex
\newacronym{IOT}{IoT}{Internet of Things}
\newacronym{PKC}{PKC}{Public-Key Cryptography}
\newacronym{RSA}{RSA}{Rivest-Shamir-Adleman}
\newacronym{ECC}{ECC}{Elliptic Curve Cryptography}
\newacronym{NIST}{NIST}{National Institute of Standards and Technology}
\newacronym{PQC}{PQC}{Post-Quantum Cryptography}
\newacronym{PKE}{PKE}{Public-Key Encryption}
\newacronym{CRYSTALS}{CRYSTALS}{Cryptographic Suite for Algebraic Lattices}
\newacronym{LWE}{LWE}{Learning With Errors}
\newacronym{RNG}{RNG}{Random Number Generator}
\newacronym{KEM}{KEM}{Key Encapsulation Mechanism}
\newacronym{MLWE}{MLWE problem}{Learning With Errors problem in module lattices}
\newacronym{CPU}{CPU}{Central Processing Unit}
\newacronym{IND-CPA}{IND-CPA}{Indistinguishability under Chosen-Plaintext Attack}
\newacronym{IND-CCA2}{IND-CCA2}{Indistinguishability under adaptive chosen ciphertext attack }
\newacronym{XOF}{XOF}{Extendable Output Function}
\newacronym{KDF}{KDF}{Key Derivation Function}
\newacronym{CCA}{CCA}{Chosen-Ciphertext Attack}
\newacronym{RTOS}{RTOS}{Real-Time Operating System}
\newacronym{AES}{AES}{Advanced Encryption Standard}
\newacronym{SHA}{SHA}{Secure Hash Algorithm}
\newacronym{ESP-IDF}{ESP-IDF}{Espressif IoT Development Framework}
\newacronym{API}{API}{Application Programming Interfaces}
\newacronym{NTT}{NTT}{Number Theoretic Transform}
\newacronym{PRF}{PRF}{Pseudo-Random Function}

%% file: lib/values.tex
\newcommand{\scenOneKeygenCycle}{2,439,083 }
\newcommand{\scenOneKeygenRunTime}{15.24 }
\newcommand{\scenOneKeygenSpeedUp}{1x }
\newcommand{\scenOneEncCycle}{2,736,256 }
\newcommand{\scenOneEncRunTime}{17.10 }
\newcommand{\scenOneEncSpeedUp}{1x }
\newcommand{\scenOneDecCycle}{2,970,640 }
\newcommand{\scenOneDecRunTime}{18.57 }
\newcommand{\scenOneDecSpeedUp}{1x }

\newcommand{\scenTwoKeygenCycle}{2,007,689 }
\newcommand{\scenTwoKeygenRunTime}{12.55 }
\newcommand{\scenTwoKeygenSpeedUp}{1.21x }
\newcommand{\scenTwoEncCycle}{2,243,652 }
\newcommand{\scenTwoEncRunTime}{14.02}
\newcommand{\scenTwoEncSpeedUp}{1.22x }
\newcommand{\scenTwoDecCycle}{2,471,286 }
\newcommand{\scenTwoDecRunTime}{15.45 }
\newcommand{\scenTwoDecSpeedUp}{1.20x }

\newcommand{\scenThreeKeygenCycle}{1,414,389 }
\newcommand{\scenThreeKeygenRunTime}{8.84 }
\newcommand{\scenThreeKeygenSpeedUp}{1.72x }
\newcommand{\scenThreeEncCycle}{1,490,784 }
\newcommand{\scenThreeEncRunTime}{9.32 }
\newcommand{\scenThreeEncSpeedUp}{1.84x }
\newcommand{\scenThreeDecCycle}{1,756,638 }
\newcommand{\scenThreeDecRunTime}{10.98 }
\newcommand{\scenThreeDecSpeedUp}{1.69x }

\newcommand{\indcpaKeygenSingleCycle}{2,248,841 }
\newcommand{\indcpaKeygenSingleRunTime}{14.06 }
\newcommand{\indcpaKeygenSingleSpeedUp}{1x }
\newcommand{\indcpaKeygenDualCycle}{1,818,840 }
\newcommand{\indcpaKeygenDualRunTime}{11.37 }
\newcommand{\indcpaKeygenDualSpeedUp}{1.24x }

\newcommand{\indcpaEncSingleCycle}{2,011,697 }
\newcommand{\indcpaEncSingleRunTime}{12.57 }
\newcommand{\indcpaEncSingleSpeedUp}{1x }
\newcommand{\indcpaEncDualCycle}{1,513,054 }
\newcommand{\indcpaEncDualRunTime}{9.46 }
\newcommand{\indcpaEncDualSpeedUp}{1.33x }

\newcommand{\indcpaDecSingleCycle}{310,357 }
\newcommand{\indcpaDecSingleRunTime}{1.94 }
\newcommand{\indcpaDecSingleSpeedUp}{1x }
\newcommand{\indcpaDecDualCycle}{353,378 }
\newcommand{\indcpaDecDualRunTime}{2.21 }
\newcommand{\indcpaDecDualSpeedUp}{0.88x }

\newcommand{\shaTfsNoAccCycle}{106,465 }
\newcommand{\shaTfsNoAccRunTime}{0.67 }
\newcommand{\shaTfsNoAccSpeedUp}{1x }
\newcommand{\shaTfsAccCycle}{10,197 }
\newcommand{\shaTfsAccRunTime}{0.06 }
\newcommand{\shaTfsAccSpeedUp}{10.44x }

\newcommand{\shaFotNoAccCycle}{474,501 }
\newcommand{\shaFotNoAccRunTime}{2.97 }
\newcommand{\shaFotNoAccSpeedUp}{1x }
\newcommand{\shaFotAccCycle}{77,803 }
\newcommand{\shaFotAccRunTime}{0.49 }
\newcommand{\shaFotAccSpeedUp}{6.1x }

\newcommand{\aesNoAccCycle}{77,233 }
\newcommand{\aesNoAccRunTime}{0.48 }
\newcommand{\aesNoAccSpeedUp}{1x }
\newcommand{\aesAccCycle}{8,005 }
\newcommand{\aesAccRunTime}{0.05 }
\newcommand{\aesAccSpeedUp}{9.65x }